\documentclass[preprint,showpacs,prl]{revtex4}

\usepackage{graphicx}

\newcommand {\nc} {\newcommand}
\nc {\beq} {\begin{eqnarray}}
\nc {\eeq} {\end{eqnarray}}
\nc {\eeqn} [1] {\label{#1} \end{eqnarray}}
\nc {\eol} {\nonumber \\}
\nc {\eoln} [1] {\label{#1} \\}
\nc {\bne} {b_\mathrm{ne}}

\begin{document}

\title{Neutron Charge Radius Deduced from Interferometric 
Bragg Reflection Technique}

\author{J.-M.\ Sparenberg\footnote{on leave from
Universit\'e Libre de Bruxelles, PNTPM-CP 229,
Campus de la Plaine, B-1050 Brussels, Belgium.} and H.\ Leeb}
\affiliation{Atominstitut der \"Osterreichischen Universit\"aten,
Technische Universit\"at Wien, \\
Wiedner Hauptstra{\ss}e 8-10, A-1040 Vienna, Austria}
\date{\today}
\begin{abstract}
The possibility of the determination of the neutron mean square charge 
radius from high-precision thermal-neutron measurements 
of the nuclear scattering length and 
of the scattering amplitudes of Bragg reflections is considered. 
Making use of the same interferometric technique as Shull in 1968, 
the scattering amplitudes of about eight higher-order Bragg reflections 
in silicon could be measured without contamination problem. 
This would provide a value
of the neutron charge radius as precise as the disagreeing Argonne-Garching and
Dubna values, as well as a Debye-Waller factor of silicon ten times 
more precise than presently available.
\end{abstract}


\maketitle


The so-called mean square charge radius of the neutron 
$\langle r_\mathrm{n}^2 \rangle $ is an important structure constant 
which reflects the internal charge structure of the neutron. 
In terms of the generally used form factors, 
$\langle r_\mathrm{n}^2\rangle $ is related to the derivative 
of the  Sachs form factor 
$G_\mathrm{E}^\mathrm{n}$ at vanishing transferred momentum $Q$,
\beq
\langle r_\mathrm{n}^2\rangle = - \frac{1}{6} 
\left. \frac{dG_\mathrm{E}^\mathrm{n}}{dQ^2}\right|_{Q=0}
\, .
\eeq 
Low-energy neutron-atom scattering experiments have been found useful 
\cite{sears:86,alexandrov:92} to determine $\langle r_\mathrm{n}^2 \rangle $ 
via high-precision measurements of the neutron-electron scattering 
length $\bne$,
\beq
\bne=\frac{1}{3} \frac{\alpha\, m_\mathrm{n} c^2}{\hbar\, c} 
\langle r_\mathrm{n}^2 \rangle
\, ,
\eeq
where $m_\mathrm{n}$ is the mass of the neutron. Because of the fundamental
importance of these quantities for our understanding of the nucleon 
many ambitious experiments have been performed in the last decades
(see e.g.\ \cite{kopecky:97} and references therein).

Despite these efforts the determination of $\bne$ is still 
unsatisfactory because there exist two sets of results which
differ more than three standard deviations from each other. In this 
paper we propose an independent method based on high-precision measurements 
of neutron Bragg reflections on silicon. Assuming for all higher-order 
reflections a precision similar to that achieved by Shull \cite{shull:68} 
one can solve the discrepancy in the values of $\bne$.  

In a first theoretical estimate, $\bne$ is expected to be given by the 
Foldy scattering length \cite{foldy:58},
\beq
\bne^\mathrm{theory}=-1.467971(4) \times 10^{-3}\,\mbox{fm},
\eeqn{bth} 
corresponding to the well-known value of the anomalous magnetic moment of
the neutron \cite{rauch:00}. The actual $\bne$ value can be deduced from
low-energy neutron-atom scattering, for which the
``scattering length'' reads \cite{sears:86}
\beq
b(Q) & = & b_\mathrm{nuclear} \underbrace{-\bne Z [1-f(Q)]}_\mathrm{electrostatic} \label{b} \\
 & = & b_\mathrm{nucleus} + \underbrace{\bne Z f(Q)}_\mathrm{electrons},
\eeq
where $b_\mathrm{nuclear}$ is the nuclear-interaction scattering length,
$b_\mathrm{nucleus}$ is the neutron-nucleus scattering length 
(nuclear and electrostatic interactions),
$Z$ is the atomic number and $f(Q)$ is the atomic form factor
normalized to the forward direction: $f(0)=1$.
In Eq.\ (\ref{b}), the $Q$-dependent electrostatic term 
is typically three orders of magnitude smaller than 
the constant nuclear term, 
hence 
it is reasonable to speak of ``scattering length'' despite the $Q$ dependence.
To get a sufficiently high sensitivity to this term scattering processes
at $Q\neq 0$ (hence $0<f(Q)<1$) and high $Z$ should be chosen.

The best experimental values of $\bne$ (see e.g.\ Table I in 
\cite{leeb:93b}) can be grouped into two sets.
Those
deduced from thermal-neutron angular scattering on noble gases \cite{khron:73} 
and from thermal- and epithermal-neutron transmission on liquid lead and
bismuth \cite{koester:95,kopecky:97} agree on the value 
$\bne=-1.31(3)\times 10^{-3}$ fm (Argonne-Garching).
The second set of $\bne$ values obtained from another transmission 
experiment on Bi
\cite{alexandrov:86} and a measurement of thermal-neutron Bragg reflections on 
tungsten monocrystals \cite{alexandrov:75} are centered about 
$\bne=-1.59(4)\times 10^{-3}$ fm (Dubna). There is a clear discrepancy because
the $\bne$ values of the two sets differ more than three standard deviations
from each other. 
In addition they differ about $10\% $ from $\bne^\mathrm{theory}$, 
one being smaller than $\bne^\mathrm{theory}$, the other larger.
While the difference in the $\bne$ values obtained from the Bi-transmission
experiments \cite{kopecky:97,alexandrov:86} is well understood
\cite{koester:95,leeb:93b,alexandrov:94},
the Bragg-reflection method of Ref.\ \cite{alexandrov:75} has never been 
revisited.
In the latter experiment, the value of $\bne$ is extracted from the slope of
$b(Q)$ as a function of $1-f(Q)$ [see Eq.\ (\ref{b})].
The fitted data are the forward value $b_\mathrm{nuclear}$, 
obtained from a Christiansen-filter experiment, and the values
for eight Bragg reflections $b(Q_{hkl})$.
The transferred momentum $Q_{hkl}$ reads
\beq
Q_{hkl}/4\pi=\sqrt{h^2+k^2+l^2}/2 a_0=\sin \theta_{hkl}/\lambda,
\eeqn{bragg}
where (hkl) are the Miller indices, 
$a_0$ is the side of the conventional cubic unit cell,
$\theta_{hkl}$ is the Bragg angle and $\lambda$ is the neutron wavelength.

For two different crystals, two very different values of $\bne$ were obtained
in Ref.\ \cite{alexandrov:75}: i.e., $b_{ne}=-1.06$ and 
$-2.2 \times 10^{-3}$ fm,
which led the authors to postulate the existence 
of an additional scattering process.
We rather consider that this discrepancy reveals the systematic 
uncertainty of the experiment, 
where $b(Q_{hkl})$ is deduced from an integral-intensity measurement 
of the Bragg peak.
This systematic uncertainty has probably been underestimated in Ref.\
\cite{alexandrov:75} for different reasons:
(i) extinction is neglected, (ii) the uncertainty on the temperature factor
(which is shown below to be of crucial importance) is neglected.
Consequently,
the accuracies obtained for $b(Q_{hkl})$ (typically 0.01 fm) seem unrealistic
as compared with accuracies generally obtained by this kind of 
measurement (typically 0.1 fm) \cite{sears:89}.
A hint to our conjecture is also given by 
the strong disagreement between other results \cite{alexandrov:70}
obtained {\em inter alia} 
with an integral-intensity measurement on monocrystals,
and the Christiansen-filter results of Ref.\ \cite{knopf:87} 
for W isotopes.

Let us now consider the method proposed by Shull in Ref.\ \cite{shull:68},
which also aims at measuring $b(Q_{hkl})$ 
for Bragg reflections on monocrystals.
We refer to Ref.\ \cite{shull:68} for details 
about the experimental setup.
A collimated full-spectrum incident beam is Bragg reflected 
on a monocrystal blade (typical thickness $t=1$ cm) 
in Laue transmission geometry.
The wavelength selection is made through the Bragg condition (\ref{bragg}),
the crystal and the detector being moved simultaneously 
to form angles $\theta$ and $2 \theta$ respectively 
with the incident-beam direction.
As shown in the dynamical theory of Bragg reflection 
for neutrons \cite{sears:89,rauch:00}, 
the reflected beam inside the crystal consists of two coherent waves,
the so-called {\em Pendell\"osung} (pendulum solutions),
interfering with each other and generating Pendell\"osung oscillations.
An entrance and a scanning gadolinium slits (width 0.13 mm) are placed on 
both faces of the crystal to extract these fringes.
Making use of the symmetry of the interference pattern a very precise alignment
of the scanning slit with respect to the entrance slit is possible.
The intensity at the center of the interference pattern is then measured as
a function of the wavelength and compared with the theoretical expression.
This expression may be obtained either by the Kato 
spherical-wave dynamical theory \cite{shull:72} 
or by solving the Takagi-Taupin equations \cite{rauch:00}.
For a perfectly-flat crystal and with infinitely-narrow slits,
this intensity reads
\beq
I_{hkl}(\lambda) \propto I_0(\lambda) \lambda^2 |F_{hkl}|^2
J_0^2\left(\frac{t |F_{hkl}|}{a_0^3 \cos \theta_{hkl}(\lambda)} \lambda \right),
\eeqn{I}
where $I_0$ is the incident intensity, $J_0$ is a Bessel function
and $|F_{hkl}|$ is the unit-cell structure factor of the crystal.
This expression has to be slightly modified to take into account the finite
curvature of the crystal, which adds the curvature radius $R$ 
(typically 20 km) as a parameter to be fitted to the data,
and to take into account the finite opening of the slits \cite{shull:72}.
Typically, forty fringes are observed for neutrons ranging the full spectrum
of a thermal reactor.
The structure factor can then be deduced with high precision 
from the period of these fringes.
Shull's method is thus in essence an interferometric method,
which explains why it provides very high accuracies 
(contrary to the integral-intensity method).

\begin{figure}
\begin{center}
\scalebox{0.47}{\includegraphics{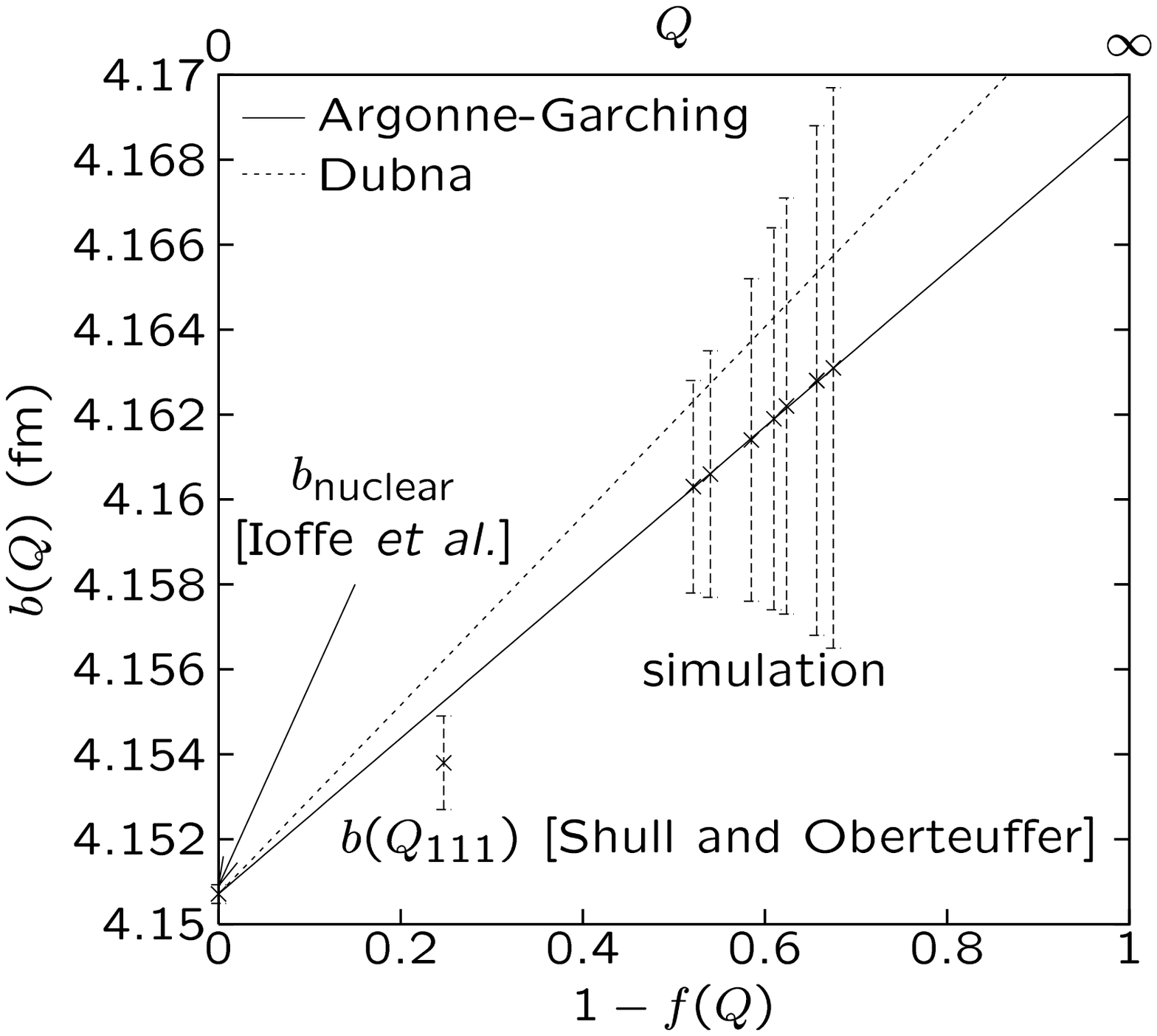}}
\caption{\label{b_Si_sim} 
Experimental $b_\mathrm{nuclear}$ of Eq.\ (\ref{bnuc})
and $b(Q_{111})$ of Eq.\ (\ref{b111}) compared
with the theoretical curves (\ref{b}) for $b_\mathrm{nuclear}=4.1507$
fm, $\bne=-1.31$ (Argonne-Garching) and
$-1.59\times 10^{-3}$ fm (Dubna).
The simulated points correspond to the reflections of Table \ref{refl};
they are calculated with the Argonne-Garching $\bne$ value,
with error bars only due to the uncertainty 
on the temperature factor (\ref{B}).}
\end{center}
\end{figure}

For a diamond-structure crystal like silicon, 
the unit cell consists of four elementary cells with two atoms.
The corresponding structure factor reads \cite{sears:89}
\beq
F_{hkl}=4\times
\left(1+i^{h+k+l}\right)\times b_\mathrm{meas}\left(Q_{hkl}\right),
\eeqn{Fhkl}
where $b_\mathrm{meas}(Q)$ is related to $b(Q)$ 
through the Debye-Waller temperature factor \cite{sears:89}
\beq
b_\mathrm{meas}(Q)=b(Q)\times \exp{\left[-B (Q/4\pi)^2\right]}.
\eeqn{bmeas}
To get a precise $b(Q_{hkl})$ value, as needed to estimate $\bne$,
one needs precise values of both the structure factor $F_{hkl}$ and the 
temperature parameter $B$.
For instance, for the (111) reflection on Si ($Z=14$, $a_0=5.43072$ \AA),
Shull's method provides \cite{shull:72} 
$b_\mathrm{meas}(Q_{111})=4.1053(8)$ fm.
Using the Debye-Waller factor obtained by measuring
X-ray Pendell\"osung fringes at room temperature \cite{aldred:73}
\beq
B=0.4613(27)\, \mbox{\AA}^2,
\eeqn{B}
one gets
\beq
b(Q_{111})=4.1538(11)\, \mbox{fm,} 
\eeqn{b111}
where three additional units in the error are due to the error on $B$.
Equation (\ref{bmeas}) shows that this additional uncertainty increases 
exponentially with $Q^2$ (see Fig.\ \ref{b_Si_sim}).

\begin{figure}
\begin{center}
\scalebox{0.49}{\includegraphics{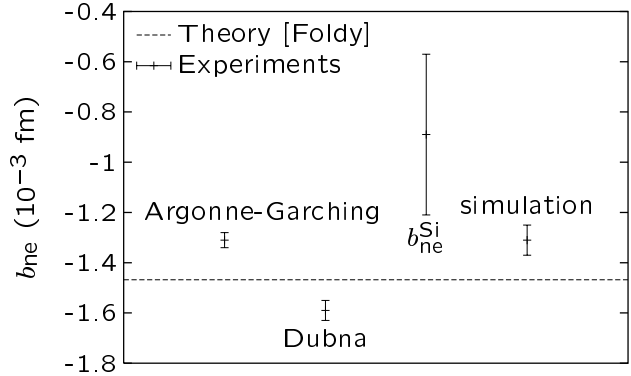}}
\caption{\label{bne} 
Comparison of the theoretical $\bne$ value (\ref{bth})
with the Argonne-Garching and Dubna experimental values
and with the result (\ref{bneSi}).
The simulated point has the Argonne-Garching value with the error bar
of Eq.\ (\ref{errors}).}
\end{center}
\end{figure}

Following the approach of Ref.\ \cite{shull:72},
we can deduce $b_\mathrm{nuclear}$ from value (\ref{b111}) 
with Eq.\ (\ref{b}), using the Argonne-Garching $\bne$ and 
the atomic form factor \cite{kromer:68}:
$f(Q_{111})=0.7526$ 
(a precise value of $f(Q)$ is actually not required \cite{sears:86}).
This yields $b_\mathrm{nuclear}=4.1495(11)$ fm 
(which slightly differs from the value of Ref.\ \cite{shull:72} because
of the more recent temperature factor),
in agreement with the more precise value obtained recently 
by non-dispersive-sweep neutron interferometry \cite{ioffe:98}
\beq
b_\mathrm{nuclear}=4.1507(2)\, \mbox{fm}.
\eeqn{bnuc}
Conversely, $\bne$ can be deduced from values (\ref{b111}) and (\ref{bnuc})
with Eq.\ (\ref{b}), which provides (see Fig.\ \ref{b_Si_sim})
\beq
\bne^\mathrm{Si}=-0.89(32)\times10^{-3}\, \mbox{fm.}
\eeqn{bneSi}
This value agrees with the Argonne-Garching value but does not
exclude the Dubna value (see Fig.\ \ref{bne}).
The same method applied to germanium (diamond structure with $a_0=5.6575$ \AA)
using $b_\mathrm{nuclear}=8.1929(17)$ fm \cite{schneider:76},
$b_\mathrm{meas}(Q_{111})=8.0829(15)$ fm \cite{shull:73b},
$B=0.57(1)$ \AA$^2$ \cite{butt:88} and $f(Q_{111})=0.8542$ \cite{kromer:68},
leads to the value $\bne^\mathrm{Ge}=0.28(83)\times10^{-3}$ fm.
Although measurements on Ge (Z=32)
could in principle lead to a value of $\bne$ two times more precise
than on Si ($Z=14$),
this advantage is compensated by the two-times-larger scattering length,
which implies a two-times-larger absolute error on $b(Q_{hkl})$,
the relative accuracy of Shull's method being 0.02\% in both cases.
Moreover, no $b_\mathrm{nuclear}$ value for Ge of comparable precision to
(\ref{bnuc}) is available at present.
Hence we only consider Si in the following.

In Fig.\ \ref{b_Si_sim}, values (\ref{b111}) and (\ref{bnuc}) are 
represented as a function of $1-f(Q)$.
It is seen that the (111) reflection is not the optimal choice 
to calculate the slope of the line since it is close to the origin.
Higher-order reflections are more appropriate, as shown by the
simulated data.
The error bars of these simulated points are calculated with the error
$\sigma_{B}=0.0027$ \AA$^2$ on the temperature factor (\ref{B}) only, 
assuming an ideal experiment with an infinite accuracy on $b_\mathrm{meas}$.
With such large uncertainties, there is no hope to reach a high precision on
$\bne$, in particular to distinguish between the Argonne-Garching 
(solid line in Fig.\ \ref{b_Si_sim}) and Dubna (broken line) values.
However, if enough (at least two) reflections are measured,
{\em both} $B$ and $\bne$ could be deduced from the data.
The precision of the extracted $B$ and $b_{ne}$ can be improved
by increasing the number of reflections.

Let us now determine the possibly-observable reflections.
A diamond-structure crystal has a face-centered cubic lattice, 
which implies that all Miller indices have to be
either even or odd \cite{sears:89}.
Moreover, for a diamond structure, the structure factor (\ref{Fhkl}) leads to
three types of reflections: (i) forbidden when $h+k+l=2+4n$,
where $n$ is natural, (ii) weak when $h+k+l$ is odd, (iii) strong when
$h+k+l=4+4n$.
A thermal-neutron beam has typically a maximum flux for $\lambda=1.2$ {\AA}
and the accessible angular range of the setup is assumed to be  
$0\le 2\theta \le 110^\circ$.
These conditions combined with Eq.\ (\ref{bragg}) imply that
sixteen reflections between (111) and (642) could be measured 
with reasonable intensities for Si.
However, if the incident beam has a typical full spectrum 
$0.8\le \lambda \le 2.5$ {\AA},
contamination has to be taken into account.
For instance, while the (111) reflection is pure for 
$15\le 2\theta \le 45^\circ$
thanks to the absence of the (222) reflection which is forbidden
(hence the result of Refs.\ \cite{shull:68,shull:72}),
(333) and (444) would be measured on $45\le 2\theta \le 110^\circ$ and
$61\le 2\theta \le 110^\circ$ respectively.
This mixing implies
that the reflected intensity would not be described by Eq.\ (\ref{I})
and that no structure factor could easily be extracted.
Among the sixteen considered reflections, seven could be contaminated.
The remaining nine reflections are listed in Table \ref{refl}.
Three crystal blades would suffice to measure them all [for instance,
a crystal blade cut parallel to the (220) planes could be used to measure
the (111), (422), (511), (533), (711) and (551) reflections].
Among these reflections, three are strong (underlined in Table \ref{refl});
the intensities range between 0.7 and 1.7 times the intensity of the
already-measured (111) weak reflection.
Finally, Eq.\ (\ref{I}) allows to estimate the number of fringes which
would be measured for each of them.
For instance, 44 fringes would be observed for the (711) reflection 
if $t=1$ cm.

These considerations show that these higher reflections should not be more
complicated to measure than the (111) reflection already measured in Refs.\
\cite{shull:68,shull:72}.
Hence, we can reasonably assume that the precision on $b_\mathrm{meas}$
would be equal to that of Ref.\ \cite{shull:72}: 0.0008 \AA.
With this assumption, the attainable precision for $B$ and $\bne$ 
can be estimated.
For $B$, we apply the approximate linear relation 
[Eqs.\ (\ref{b}) and (\ref{bmeas})]
\beq
\ln b_\mathrm{meas}(Q)\approx \ln b_\mathrm{nuclear} - B (Q/4\pi)^2,
\eeq
while for $\bne$ we use Eq.\ (\ref{b}).
The uncertainty of a linear-fit slope can be calculated from the
abscissas of the experimental points and from their errors \cite{press:89}.
Measuring the three strong reflections would give
\beq
\sigma_B = 0.00040\,\mbox{\AA}^2,\,
\sigma_{\bne} = 0.11 \times 10^{-3} \, \mathrm{fm},
\eeq
while with the eight new reflections of Table \ref{refl}, one gets
\beq
\sigma_B = 0.00027\,\mbox{\AA}^2, \,
\sigma_{\bne} = 0.06 \times 10^{-3} \, \mathrm{fm}.
\eeqn{errors}
The temperature factor would be ten times more precise than the most
precise value (\ref{B}) and the $\bne$ value would be nearly as precise
as the Argonne-Garching and Dubna values (see Fig.\ \ref{bne}).
Since the present method is similar to the Dubna method on W,
such a new result would certainly be of great help 
to decide the $\bne$ discrepancy.  

In conclusion, the measurement of Pendell\"osung fringes in Bragg reflections 
is a very powerful technique.
This is known for X-rays \cite{aldred:73,saka:86} but generally
ignored for neutrons.
By confirming Shull's pioneer conclusion \cite{shull:68}
\begin{quote} 
``\dots the high sensitivity of the fringe positions \dots can be
exploited \dots [for] example \dots 
in assessing the neutron-electron interaction strength\dots''
\end{quote}
the present work opens new prospects for the field of
{\em Neutron Pendell\"osung Interferometry}.

We acknowledge useful discussions
with H.\ Rauch, H.\ Kaiser and E.\ Jericha.
J.-M.\ S.\ is supported by the European TMR program ERB-FMRX-CT96-0057
on perfect-crystal neutron optics and by 
the National Fund for Scientific Research (Belgium). 
He thanks the members of the Atominstitut for their hospitality.
%

\newpage

\newpage
%
\begin{table}
\begin{center}
\caption{\label{refl} 
Weak and strong (underlined) Bragg reflections
which could be measured with a thermal-neutron reactor
without contamination problem. (hkl) are the Miller indices, 
$f(Q_{hkl})$ is the atomic form factor \protect\cite{kromer:68},
$\lambda$ the neutron-wavelength interval,
$2 \theta$ the corresponding angles between the incident and reflected beams.
$|F_{hkl}|^2$ is the squared structure factor 
which provides the reflection intensity through Eq.\ (\ref{Fhkl}).}
\begin{tabular}{|c|c|c|c|c|}
\hline
$(hkl)$ & $f(Q_{hkl})$ & $\lambda$ (\AA) & $2 \theta(^\circ)$ & 
$|F_{hkl}|^2$ (\AA$^2$) \\
\hline
(111) & 0.7526 & 0.8-2.5 & 15-47 & 540 \\
(\underline{422}) & 0.4788 & 0.8-1.8 & 42-110 & \underline{918} \\
(511) & 0.4600 & 0.8-1.7 & 45-110 & 448 \\
(531) & 0.4150 & 0.8-1.5 & 52-110 & 421 \\
(\underline{620}) & 0.3902 & 0.8-1.4 & 56-110 & \underline{811} \\
(533) & 0.3764 & 0.8-1.4 & 58-110 & 396 \\
(551) & 0.3432 & 0.8-1.2 & 63-110 & 372 \\
(711) & 0.3432 & 0.8-1.2 & 63-110 & 372 \\
(\underline{642}) & 0.3249 & 0.8-1.2 & 67-112 & \underline{715} \\           
\hline 
\end{tabular}
\end{center}
\end{table}

%
\end{document}